\def\erfc{\mathop{\rm erfc}}
\def\erf{\mathop{\rm erf}}

\makeatother
\documentclass[aps,pre,floats,twocolumn,superscriptaddress,%
  amssymb,amsmath,showpacs]{revtex4}
\usepackage{amssymb}

\usepackage{epsf}
\usepackage{subfigure}
\usepackage{graphicx}
\usepackage{color}
\usepackage[hypertex]{hyperref}
\begin{document}
\title{Attachment and detachment rate distributions in deep-bed filtration}
\author{Hsiang-Ku Lin}
\affiliation{Department of Physics and Astronomy, University of California,
  Riverside, California 92521, USA}
\author{Leonid P. Pryadko}
\affiliation{Department of Physics and Astronomy, University of California,
  Riverside, California 92521, USA}
  \author{Sharon Walker}
  \affiliation{Department of Chemical and Environmental Engineering,  
University of California,
   Riverside, California 92521, USA}

\author{Roya Zandi}
\affiliation{Department of Physics and Astronomy, University of California,
  Riverside, California 92521, USA}
\date\today
\begin{abstract}
  We study the transport and deposition dynamics of colloids 
  in saturated porous media under unfavorable filtering conditions.
  As an alternative to traditional convection-diffusion or more
  detailed numerical models, we consider a mean-field description in which
  the attachment and detachment processes are characterized by an entire
  spectrum of rate constants, ranging from shallow traps which mostly
  account for hydrodynamic dispersivity, all the way to the permanent
  traps associated with physical straining.  The model has an
  analytical solution which allows analysis of its properties
  including the long time asymptotic behavior and the profile of the
  deposition curves. Furthermore, the model gives rise to a filtering
  front whose structure, stability and propagation velocity are
  examined.  Based on these results, we propose an experimental
  protocol to determine the parameters of the model.
\end{abstract}
 \maketitle

\section{Introduction}
Many processes in biological systems as well as in the chemical and petroleum
industry involve the transport and filtration of particles in porous media with
which they interact through various
forces\cite{Indakm-1987,Bloomfield-1999,Luhrmann-1999,Harvey-1991}. These
interactions often result in particle adsorption and/or entrapment by the
medium.  Examples include filtration in the respiratory system, groundwater
transport, in situ bioremediation, passage of white blood cells in brain blood
vessels in the presence of jam-1 proteins, passage of viral particles in
granular media, separation of species in chromatography, and gel permeation.
The particle-medium interactions in these systems are not always optimal for
particle retention.  For example, the passage of groundwater through soil
often happens under chemically unfavorable conditions, and as a result many
captured particles (e.g., viruses and bacteria) may be released back to the
solution. While filtration under favorable conditions has been studied and
modeled extensively\cite{Herzig-1970,Tien-1979,Chiang-1985,Vigneswaran-1986,%
  Vigneswaran-1988,Ghidaglia-1991,Tobiason-1994,Lee-Koplik-1996,%
  Vigneswaran-1989,Jegatheesan-2005}, we are just beginning to understand the
process occurring under unfavorable conditions.

Several models have been developed to describe the kinetics of particle
filtration under unfavorable conditions. The most commonly used ones are, in
essence, phenomenological mean-field models based on the convection-diffusion
equation (CDE) [see Eq.~(\ref{eq:convection-diffusion}) and
Sec.~\ref{sec:background}].  Typically, one models the dynamics of free
particles in terms of the average drift velocity $v$ and the hydrodynamic
dispersivity $\lambda$, while the net particle deposition rate $r_d$
accounting for particle attachment and detachment at trapping sites is a
few-parameter function of local densities of free and trapped particles.  For
given filtering conditions, the parameters $\lambda$ and $v$ can be determined
from a separate experiment with a tracer, while the coefficients of the
function $r_d$ can be obtained by fitting Eq.~(\ref{eq:convection-diffusion})
to the breakthrough curves.

Despite their attractive simplicity, it is widely accepted now that the
phenomenological models at the mean-field level have significant problems.
First, the depth dependent deposition curves for viruses and bacteria are
often much steeper than it would be expected if the deposition rates were
uniform throughout the substrate\cite{Abinger-1994,Baygents-1998,Simoni-1998,%
  Bolster-1999,Bolster-2000,Redman-2001,Redman-2001B,Tufenkji-2003}.  This was
commonly compensated by introducing the depth-dependent deposition rates.  The
problem was brought to light in Ref.~\cite{li-2004}, where it was demonstrated
that the steeper-than-expected deposition rates under unfavorable filtering
conditions also exist for inert colloids.

Second, Bradford et al. \cite{Bradford-2002,Bradford-2003} pointed
out that the usual mean-field models based on the CDE, accounting for
dynamic dispersivity and attachment and detachment phenomena, cannot explain the
shape of both the breakthrough curves and the subsequent filter flushing.  In
these experiments some particles were retained in the medium, and
the authors argued for the need to include the straining (permanent capture of
colloids) in the model.  Even so, these models may still be insufficient to fit the
experiments \cite{Yoon-2006}.

More elaborate models to describe deep-bed filtration have been
proposed in
Refs.~\cite{Redner-2000,Hwang-2001,Kim-2006,shapiro-2007}. These
models go beyond the mean-field description by simulating subsequent
filter layers as a collection of multiply connected pipes with a wide
distribution of radii, which results in a variation in flow speed and
also of the attachment and detachment rates (even straining in some
cases).  The disadvantage of these models is that they are essentially
computer based: it is difficult to gain an understanding of the
qualitative properties of the solutions, without
extensive simulations.  Furthermore, the simulation results suffer
from statistical uncertainties.  

In the present work, we develop a minimalist mean-field model to
investigate filtering under unfavorable conditions.  The model
accounts for both a convective flow and the primary attachment and detachment
processes.  Unlike the previous mean-field models of filtration, our
model contains attachment sites (traps) with different detachment
rates $B_i$ [see Eq.~(\ref{eq:nonlin-trap})], which allows an accurate
modeling of the filtration dynamics over long-time periods for a broad range
of inlet concentrations.  Yet, the model admits exact analytical solutions
for the profiles of the deposition and breakthrough curves which
permit us to understand qualitatively the effect of the corresponding
parameters and design a protocol for extracting them from experiment.

One of the advantages of our model is that the ``shallow'' short-lived traps
represent the same effect as hydrodynamical dispersivity without generating
unphysically fast moving particles or requiring an additional boundary
condition at the inlet of the filter.  The ``deep'' long-lived traps allow to
correctly simulate long-time asymptotics of the released colloids in the
effluent during a washout stage.  The traps with intermediate detachment rates
determine the most prominent features of breakthrough curves.  The effect of
every trap kind is to decrease the apparent drift velocity.  As
attachment and detachment rate constants depend on colloid size, we can also
account for the apparent acceleration of larger particles without any
microscopic description as in Ref.~\cite{Scheibe-2003}.  The particle-size
distribution can be also used to analyze the steeper deposition profiles near
the inlet of the filter \cite{Baygents-1998,Simoni-1998,Tufenkji-2003,li-2004}.

The paper is organized as follows.  In Sec.~\ref{sec:background}, we
give a brief overview of colloid-transport experiments, CDE models,
and their analytical solutions in simple cases.  The linearized
multirate convection-only filtration model is introduced in
Sec.~\ref{sec:linearized}.  The model is characterized by a discrete
or continuous trap-release-rate distribution; it is generally solved
in quadratures, and completely in several special cases.  The results
support our argument that the hydrodynamic dispersivity can be traded
for shallow traps.  This serves as a basis for the exact solution of
the full mean-field model for filtration under unfavorable conditions
introduced in Sec.~\ref{sec:full}, where we show that a large class of
such models can be mapped exactly back to the linearized ones and
analyze their solutions, as well as the propagation velocity,
structure, and stability of the filtering front.  We suggest an
experimental protocol to fit the parameters of the model in
Sec.~\ref{sec:experiment} and give our conclusions in
Sec.~\ref{sec:conclusions}.

\section{Background}
\label{sec:background}
\subsection{Overview of colloid transport experiments}
\begin{figure}[htp]
  \includegraphics[width=.18\textwidth]{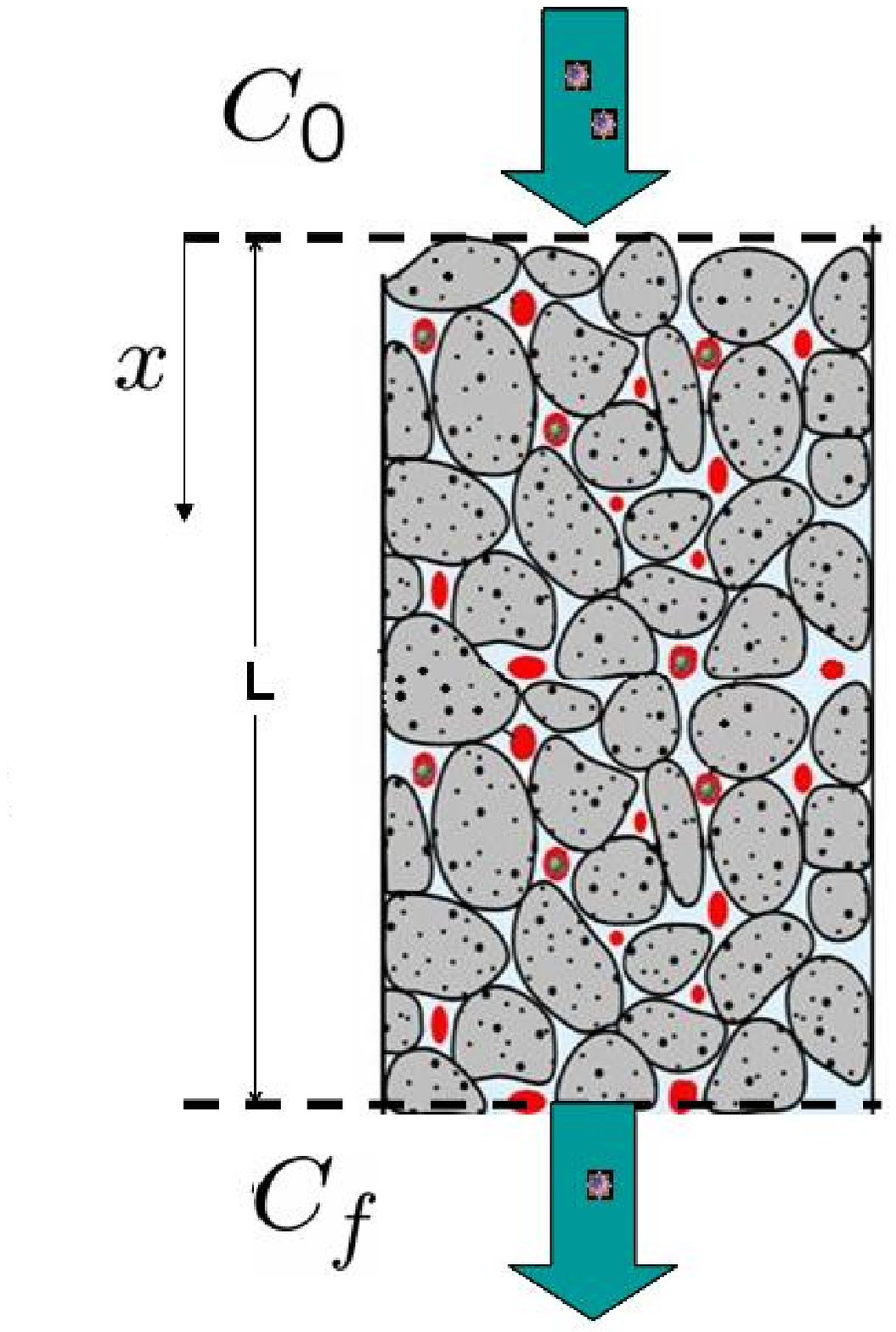}
  \caption{Schematic of experimental setup in the colloid-transport studies.}
  \label{fig:setup}
\end{figure}

A typical setup of a colloid-transport experiment is shown in
Fig.~\ref{fig:setup}.  A cylindrical column packed with sand or other
filtering material is saturated with water running from top to bottom until
the single-phase state (no trapped air bubbles) is achieved.  At the end of
this stage, colloidal particles are added to the incoming stream of water with
both the concentration of the suspended particles and the flow rate kept
constant over time $T$.  This is sometimes followed by a filter washout stage
in which clean water is pumped through the filter.  The filtration processes
are characterized by two relevant experimental quantities: the particle
breakthrough and deposition profile curves. While breakthrough curve
represents the concentration of effluent particles at the outlet of the column
as a function of time, deposition curves illustrate the depth distribution of
concentration of the particles retained throughout the column.

\subsection{Convection-diffusion transport model}
\label{sec:convection-diffusion}
As the suspended particles move through the filtering column, each
individual colloid follows its own trajectory.  Consequently, even for
small particles that are never trapped in the filter, the passage time
through the column fluctuates.  In the case of laminar flows with
small Reynolds numbers and sufficiently small particles, which
presumably follow the local velocity lines, the passage time scales
inversely with the average flow velocity along the column $v$.  The
effects of the variation between the trajectories of particles as well
as their speeds can be approximated by the velocity-dependent
diffusion coefficient $D=\lambda v$, where $\lambda$ is the {\em
  hydrodynamic dispersivity\/} of the filtering medium.  In
comparison, the actual diffusion rate of colloids in experiments is
negligibly small.  Dispersivity is often obtained through tracer
experiments in which the motion of the particles, i.e., salt ions,
which move passively through the filter medium without being trapped,
is traced as a function of time.

Overall, the dynamics of the suspended particles along the filter can be
approximated by the mean-field CDE,
\begin{equation}
  \frac{\partial C}{\partial t}+v\frac{\partial C}{\partial x}-
    \lambda v \frac{\partial^2 
    C}{\partial x^2}=-r_d,
  \label{eq:convection-diffusion}
\end{equation}
where $C\equiv C(x,t)$ is the number of suspended particles per unit
water volume averaged over the filter cross section at a given
distance $x$ from the inlet and $r_d$ is the deposition rate which
may include both attachment and detachment processes.

\subsection{Issues with the CDE approximation}
The diffusion approximation employed in
Eq.~(\ref{eq:convection-diffusion}) has two drawbacks which could
seriously affect the resulting calculations if enough care is not used.

{\em First\/}, while the diffusion approximation works well to describe the
concentration $C(x,t)$ of suspended particles in places where $C(x,t)$ is
large, it seems to significantly overestimate the number of particles far
downstream where $C(x,t)$ is expected to be small or zero.  This is mainly due
to the fact that the diffusion process allows for infinitely fast transport,
albeit for a vanishingly small fraction of particles.  In the simple case of
tracer dynamics [$r_d=0$ in Eq.~(\ref{eq:convection-diffusion})], the general
solutions as presented in Eqs.~(\ref{eq:tracer-convolution}) and
(\ref{eq:tracer-gf}) are non-zero even at very large distances $x- vt\gg
2(\lambda v t)^{1/2}$.  While in many instances this may not be crucial, the
application of the model to, e.g., public health and water safety issues might
trigger a false alert.

{\em Second\/}, for the filtering problem one expects the concentration
$C(x,t)$ to be continuous, with the concentration downstream uniquely
determined by that of the upstream.  On the other hand,
Eq.~(\ref{eq:convection-diffusion}) contains second spatial derivative, which
requires in addition to the knowledge of $C(x,t)$ at the inlet, $x=0$, another
type of boundary condition to describe the concentration of particles along
the column. This additional boundary condition could be, e.g., the spatial
derivative $C'(x,t)$ at the inlet, $x=0$, or the outlet,
$x=L$~\cite{Lapidus-1952,Tufenkji-2003}, or the fixed value of the
concentration at the outlet.  We show below that fixing a derivative
introduces an incontrollable error.  On the other hand, we cannot introduce a
boundary condition for the function $C(x,t)$ at the outlet, $x=L$, as this is
precisely the quantity of interest to calculate.

The situation has an analogy in neutron physics\cite{neutron-diffusion}.
While neutrons propagate diffusively within a medium, they move ballistically
in vacuum.  A correct calculation of the neutron flux requires a detailed
simulation of the momentum distribution function within a few mean-free paths
from the surface separating vacuum and the medium.  In contrast to the
filtration theory, for the case of neutron scattering, where the neutron
distribution is stationary it is common to use an approximate boundary
condition in terms of a ``linear extrapolation distance'' (the inverse
logarithmic derivative of neutron density).
  
The CDE [see Eq.~(\ref{eq:convection-diffusion})] can be solved on a
semi-infinite interval ($x_{\rm max}\gg L$) with setting $C'(x,t)=0$ at $x_{\rm
  max}$ and calculating the value of $C(x,t)$ at $x=L$ as an approximation for
the concentration of effluent particles.  To illustrate this situation, we
solve Eq.~(\ref{eq:convection-diffusion}) for the case of tracer particles,
where the deposition rate is set to zero, $r_d=0$.  We consider a
semi-infinite geometry with the initial condition $C(x,0)=0$ and a given
concentration $C(0,t)$ at the inlet.  The corresponding solution is presented
in Sec.~\ref{sec:tracer}.  The spatial derivative at the boundary given in
Eq.~(\ref{eq:tracer-step-solution-bc}) is non-zero, time-dependent, and rather
large at early stages of evolution when the diffusive current near the
boundary is large.  Therefore, setting an additional boundary condition for
the derivative, e.g., $C'(0,t)=0$, is unphysical.
  
On the other hand, the problem with the boundary condition far
downstream, $C(x_{\rm max},t)=0$, $x_{\rm max}\gg L$, can be
ill-defined numerically, as this condition is automatically satisfied
to a good accuracy as long as the bulk of the colloids has not
reached the end of the interval.

\subsection{Tracer model}
\label{sec:tracer}
The simplest version of the convection-diffusion
equation [Eq.~(\ref{eq:convection-diffusion})] applies to tracer particles where the 
deposition rate is set to zero, $r_d=0$,
\begin{equation}
  \label{eq:tracer-cde}
  \frac{\partial C}{\partial t}+v\frac{\partial C}{\partial x}-  \lambda v
  \frac{\partial^2 
    C}{\partial x^2}=0.
\end{equation}

With the initial conditions, $C(x,0)=0$, the Laplace-transformed
function $\tilde C\equiv \tilde C(x,p)$ obeys the equation
\begin{equation}
  \label{eq:tracer-laplace}
  p \tilde C+v \tilde C'-\lambda v \tilde C''=0,
\end{equation}
where primes denote the spatial derivatives, $\tilde C'\equiv
  \partial_x \tilde C(x,p )$.  The solution to the above equation is
$\tilde C\propto e^{\kappa x}$, with
\begin{equation}
  \kappa_\pm =
  {1\over 2\lambda}\pm  \left({1\over
      4\lambda ^2}+{p\over 
      \lambda v }\right)^{1/2}.\label{eq:tracer-roots}
\end{equation}
At semi-infinite interval $x>0$, only the solution with negative
$\kappa=\kappa_-$ does not diverge at infinity.  Given the Laplace-transformed
concentration at the inlet, $\tilde C(0,p)$, we obtain
\begin{equation}
  \label{eq:tracer-laplace-solution}
  \tilde C(x,p)=\tilde C(0,p) \exp \left({x\over 2\lambda}-x
    \left[{1\over
        4\lambda ^2}+{p\over 
        \lambda v }\right]^{1/2}\right).
\end{equation}
The inverse Laplace transformation of the above equation is a
convolution, 
\begin{equation}
  \label{eq:tracer-convolution}
  C(x,t)=\int_0^{t} dt'\,C(0,t-t')\,g(x,t'),
\end{equation}
with the tracer Green's function (GF)
\begin{equation}
  \label{eq:tracer-gf}
  g(x,t)=
  {x\over 2 (\pi \lambda
    v)^{1/2}}
  {1\over t^{3/2}} 
  \exp\left(-{(x-v t)^2\over 4\lambda v t}\right).
\end{equation}
In the special case $C(0,t)=C_0=$const, the integration results 
\begin{equation}
  \label{eq:tracer-step-solution}
  C=\frac{C_0}{2}
 \left(1+\erf\left[\frac{t v-x}{2 ({t v \lambda
 })^{1/2}}\right]+e^{x/\lambda } \erfc\left[\frac{t v+x}{2 ({t v
 \lambda })^{1/2}}\right]\right), 
\end{equation}
where $\erfc(z)\equiv 1-\erf(z)$ is the complementary error function. 

We note that the spatial derivative of the
solution of Eq.~(\ref{eq:tracer-step-solution}) at $x=0$ is different from
zero.  Indeed,  it depends on time and is divergent at small $t$, implying an
unphysically large diffusive component of the particle current,
\begin{equation}
\label{eq:tracer-step-solution-bc}
C'(0,t)
=\frac{C_0}{2}
 \left({\erfc\left(\alpha\right)\over 2\lambda}
-{e^{-\alpha^2}\over (\pi t v
    \lambda)^{1/2}}\right),\quad \alpha^2\equiv {tv\over 4\lambda}.
\end{equation}

In the presence of the straining term, $r_d=A_0 N_0 C$ in 
Eq.~(\ref{eq:convection-diffusion}), the GF can be obtained from
Eq.~(\ref{eq:tracer-gf}) by introducing exponential decay with the
rate $A_0 N_0$, 
\begin{equation}
  g(x,t)={x\over 2 (\pi \lambda
    v)^{1/2}}\,
  {e ^{-A_0 N_0t}\over t^{3/2}} 
  \exp\left(-{(x-v t)^2\over 4\lambda v
      t}\right).
\label{eq:cde-straining-gf}
\end{equation}
Note that we wrote the straining rate as a product of the capture rate
$A_0$ by infinite-capacity ``permanent'' traps with the concentration
$N_0$ per unit volume of water.  Such a factorization is convenient
for the non-linear model presented later in Sec.~\ref{sec:full}. The
same notations are employed throughout this work for consistency.


\section{Linearized mean-field filtration model}
\label{sec:linearized}
In this section we discuss the linearized convection-only multitrap
filtration model, a variant of the multirate CDE model first proposed in
Ref.~\cite{Haggerty-1995}.  Our model is characterized by a (possibly
continuous) density of traps as a function of detachment rate [see
Eq.~(\ref{eq:ntrap-density})].  Generically, continuous trap distribution leads
to non-exponential (e.g., power-law) asymptotic forms of the concentration in
the effluent on the washout stage.

The main purpose of this section is to demonstrate that ``shallow'' traps with
large detachment rates have the same effect as the hydrodynamic dispersivity
in CDE.  In addition, the obtained exact solutions will be used in
Sec.~\ref{sec:full} as a basis for the analysis of the full non-linear
mean-field model for filtration under unfavorable conditions.

\subsection{Shallow traps as a substitute for diffusion}
To rectify the problems with the diffusion approximation noted previously, we
suggest an alternative approach for the propagation of particles
through the filtering medium.  Instead of considering the drift with
an average velocity with symmetric diffusion-like deviations
accounting for dispersion of individual trajectories, we consider the
convective motion with the maximum velocity $v$.  The random twists
and turns delaying the individual trajectories are accounted for by
introducing Poissonian traps which slow down the passage of the
majority of the particles through the column.  In the simplest case
suitable for tracer particles, the relevant kinetic equations read as
follows:
\begin{equation}
  \label{eq:1trap-convection}
  \dot C+vC'+N_1 \dot n_1=0, \quad 
  \dot n_1=A_1 C-B_1 n_1,
\end{equation}
with $n_1\equiv n_1(x,t)$ as the auxiliary variable describing the average number
of particles in a trap, $N_1$ as the number of traps per unit water volume, $A_1$
as the trapping rate, and $B_1$ as the release rate.  The particular normalization
of the coefficients is chosen to simplify the formulation of models with traps
subject to saturation in Sec.~\ref{sec:full}.

To simulate dispersivity where all time scales are inversely
proportional to propagation velocity, we must choose both $A_1$ and
$B_1$ proportional to $v$.  The corresponding parameter
  $\sigma$ in $A_1\equiv \sigma v$ has a dimension of area and can be
viewed as a trapping cross section.  The length $\ell$ in the release
rate $B_1\equiv v/\ell$ can be viewed as a characteristic size of a
stagnation region. On general grounds we expect $\sigma\propto \ell^2$
with $\ell$ on the order of the grain size.

\subsection{Single-trap model with straining.} 
\label{sec:1trap-straining}
 To illustrate how shallow traps can provide for
  dispersivity in convection-only
  models, let us construct the exact solution of
  Eq.~(\ref{eq:1trap-convection}).  In fact, it is convenient to
  consider a slightly generalized model with the addition of
  straining, 
\begin{equation}
  \label{eq:1trap-straining}
  \dot C+v C'+N_1\dot n_1=-A_0 N_0 C ,\quad 
  \dot n_1=A_1  C-B_1 n_1.
\end{equation}
With zero initial conditions   the Laplace
transformation gives for $\tilde C\equiv \tilde C(x,p)$,
\begin{equation}
  \label{eq:1trap-straining-laplace}
  \left(p +A_0N_0 +{A_1N_1 p\over p+B_1}\right)\tilde C +v \tilde C'=0.
\end{equation}
The boundary value for Laplace-transformed $C(x,t)$ at the inlet is given by
$\tilde C(0,p)$.  
With initially clean filter, $C(x,0)=n(x,0)=0$, and a given free
particle concentration $C(0,t)$ at the inlet, the solution
to the linear one-trap convection-only
model with straining [Eq.~(\ref{eq:1trap-straining})] is a convolution
of the form presented in
Eq.~(\ref{eq:tracer-convolution}) with the following
GF \cite{endnote}:
\begin{eqnarray}
  \nonumber 
  g(x,t)&=&
e^{-\beta x/v-B_1(t- x/v)}\Biggl\{
\delta\left(t-{x}/{v}\right) \\
& & +\theta(t-x/v)\frac{(A_1 N_1 B_1 x)^{1/2}}{({tv -x})^{1/2}} I_1(\zeta_t)
  \Biggr\},
  \label{eq:1trap-straining-gf}
\end{eqnarray}
where $\beta\equiv  A_0N_0+A_1N_1$ is the clean-bed trapping rate,
$\theta(z)$ is 
the Heaviside step-function, and $I_1(\zeta_t)$
is the modified Bessel function of the first kind with the argument
\begin{equation}
  \label{eq:1trap-straining-argument}
  \zeta_t\equiv {2\over v}\left[A_1N_1 B_1(tv -x) x\right]^{1/2}.
\end{equation}
The singular term with the $\delta$ function $\delta(t-x/v)$ in
Eq.~(\ref{eq:1trap-straining-gf}) represents the particles at the
leading edge which propagate freely with the maximum velocity $v$
without ever being trapped.  The corresponding weight $\exp(-\beta
x/v)$ decreases exponentially with 
the distance from the origin.

Sufficiently far from both the origin and from the leading edge, where
the argument $\zeta_t$ [Eq.~(\ref{eq:1trap-straining-argument})] of
the Bessel function is large, we can use the asymptotic form,
\begin{equation}
  \label{eq:bessel-asymptotic}
  I_1(\zeta)={1\over (2\pi\zeta)^{1/2}}
  e^\zeta\left[1+\mathcal{O}(\zeta^{-1})\right],\mbox{Re}\,\zeta>0. 
\end{equation}
Subsequently, Eq.~(\ref{eq:1trap-straining-gf}) becomes
\begin{equation}
  \label{eq:1trap-straining-gf-simplified}
  g(x,t)\approx e^{-A_0N_0x/v} {B_1\xi^{1/4}\over 2 \pi^{1/2}
  \tau^{3/4}}\exp-(\sqrt\xi-\sqrt\tau)^{2},   
\end{equation}
where $\tau\equiv B_1(t-x/v)$ is the dimensionless retarded time in units of
the release rate, and $\xi\equiv A_1 N_1 x/v$ is the dimensionless distance
from the origin in units of the trapping mean free path.

The correspondence with the
  GF in Eq.~(\ref{eq:cde-straining-gf}) for the CDE with linear straining [or
    Eq.~(\ref{eq:tracer-gf}) for the CDE tracer model in the case of
    no permanent traps, $N_0=0$] can be recovered from
  Eq.~(\ref{eq:1trap-straining-gf-simplified}) by expanding the
  square roots in the exponent around its maximum at $\xi=\tau$, or
  $x=v_0 t$, with the effective velocity $v_0=v
  B_1/(B_1 +N_1 A_1)$.  Specifically, suppressing the prefactor due to
  straining, [$N_0=0$ in
  Eq.~(\ref{eq:1trap-straining-gf-simplified})], we obtain for the
  asymptotic form of the exponent at large $t$, 
  \begin{equation}
    \label{eq:1trap-convection-gf-gaussian}
    g(x,t)\propto \exp -{(x-v_0t)^2\over 4 \lambda_0v_0t},
  \end{equation}
with the effective dispersivity coefficient
  [cf.~Eq.~(\ref{eq:tracer-gf})] 
\begin{equation}
  \label{eq:1trap-convection-Deff}
  \lambda_0 =v {N_1A_1\over (N_1A_1+B_1)^2}. 
\end{equation}
The approximation is expected to be good as long as both $x$ and $t$
are large compared to the width of the bell-shaped maximum. 

\begin{figure}[htp]
  \centering
    \includegraphics[width=.9\columnwidth]{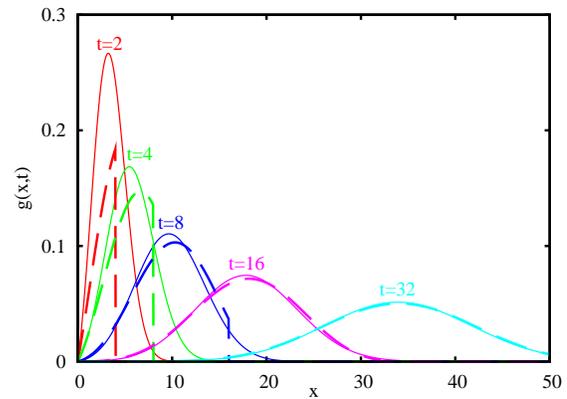}
    \caption{(Color online) Comparison of the spatial dependence of the GFs
      for the tracer model implemented as the convection-diffusion
      equation [Eq.~(\ref{eq:convection-diffusion})] with $r_d=0$
      (solid lines) and the single-trap convection model
      [Eq.~(\ref{eq:1trap-convection})] (dashed lines).  Specifically,
      we plot Eq.~(\protect\ref{eq:tracer-gf}) and the regular part of
      Eq.~(\ref{eq:1trap-straining-gf}) with $N_0=0$, using identical
      values of $v=v_0=1$ and $\lambda=\lambda_0=1$ and the release
      rate $B_1=1/2$ (half the maximum value at these parameters) at
      $t=2$, $4$, $8$, $16$, $32$.  Once the maximum is sufficiently
      far from the origin, the two GFs are virtually identical (see
      Sec.~\ref{sec:1trap-straining}).}
  \label{fig:comp}
\end{figure}

The actual shapes of the corresponding GFs, Eqs.~(\ref{eq:tracer-gf})
and (\ref{eq:1trap-straining-gf}) in the absence of permanent traps,
$N_0=0$, are compared in Fig.~\ref{fig:comp}.  While the shape
differences are substantial at small $t$, they disappear almost
entirely at later times.

\subsection{Multitrap convection-only model}
Even though the solutions of the single-trap model correspond to those
of the CDE [Eq.~(\ref{eq:tracer-cde})], the
model presented in Eq.~(\ref{eq:1trap-straining}) is clearly too simple to
accurately describe filtration under conditions where trapped particles
can be subsequently released. At the very least, in addition to
straining and ``shallow'' traps that account for the dispersivity,
describing the experiments\cite{Bradford-2002,Bradford-2003} requires
another set of ``deeper'' traps with a smaller release rate.
  
More generally, consider a linear model with $m$ types of traps
differing by the rate coefficients $A_i$, $B_i$,
\begin{equation}
  \label{eq:ntrap-convection}
  \dot C+v C'+\sum_{i=1}^m N_i \dot n_i=0,\quad 
  \dot n_i=A_i C-B_i n_i.
\end{equation}
The corresponding solution can be obtained in quadratures in terms of
the Laplace transformation.  With the initial condition,
$C(x,0)=n_i(x,0)=0$ and a given time-dependent concentration at the
inlet, $C(0,t)=C_0(t)$, the result for $C(x,t)$ is a convolution of
the form presented in Eq.~(\ref{eq:tracer-convolution}) with the GF given by 
the inverse Laplace transformation formula,
\begin{equation}
g(x,t)= \int_{c-i\infty }^{c+i\infty} {dp\over 2\pi i}
e^{\,\textstyle
  p\,[t-x/v-x\Sigma(p)/v]},\label{eq:ntrap-convection-gf}
\end{equation}
with the response function 
\begin{equation}
  \label{eq:ntrap-convection-response}
  \Sigma(p)\equiv \sum_{i=1}^m {A_i N_i\over p+B_i}=\int {dB\,
  \rho(B)\over p+B}.
\end{equation}
Here we introduced the effective density of traps, 
\begin{equation}
  \rho(B)\equiv \sum_{i=1}^m A_i N_i\delta(B-B_i), 
  \label{eq:ntrap-density}
\end{equation}
corresponding to various release rates.

The general structure of the concentration profile can be read off
directly from Eq.~(\ref{eq:ntrap-convection-gf}).  It gives zero for
$t<x/v$, consistent with the fact that $v$ is the maximum propagation
velocity in Eq.~(\ref{eq:ntrap-convection}).  The structure of
the leading-edge singularity (the amplitude of the $\delta$ function
due to particles which never got trapped) is determined by the
large-$p$ asymptotics of the integrand in
Eq.~(\ref{eq:ntrap-convection-gf}).  Specifically,
GF~(\ref{eq:ntrap-convection-gf}) can be written as
\begin{equation}
  \label{eq:ntrap-convection-gf-leading}
  g(x,t)=e^{-\beta x/v}\delta(t-x/v)+\theta(t-x/v) g_{\rm
  reg}(x,t),
\end{equation}
where 
$\beta=\lim_{p\to\infty}p\Sigma(p)=\sum_i N_i A_i$
[cf.~Eq.~(\ref{eq:1trap-straining-gf})] is the clean-bed trapping
rate, and $g_{\rm reg}$ is 
the non-singular part of the GF.

Similarly, the structure of the diffusion-like peak of the GF away
from both the origin and the leading edge is determined by the saddle
point of the integrand in Eq.~(\ref{eq:ntrap-convection-gf})
at small $p$.  Assuming the expansion $\Sigma(p)=
\Sigma(0)-\Sigma_1 p+\mathcal{O}(p^2)$ and 
evaluating the resulting Gaussian integral around the
  saddle point at 
\begin{equation}
  \label{eq:ntrap-convection-saddle}
  p_{\star}\approx{t-x/v_0\over 2 x \Sigma_1/ v},\quad
  v_0\equiv{v\over 1+\Sigma(0)},
\end{equation}
we obtain 
\begin{equation}
  \label{eq:ntrap-convection-gf-max}
  g(x,t)\approx {1\over 2 (\pi  \Sigma_1 x/v)^{1/2}} e^{-(t-x/v_0)^2/
  (4\Sigma_1 x/v)}.
\end{equation}
The exponent near the maximum can be approximately rewritten in the
form of that in Eq.~(\ref{eq:tracer-gf}), with the effective
dispersivity
\begin{equation}
  \label{eq:ntrap-eff-params}
  \lambda_0=  {v_0^2\over v}\Sigma_1
  =  
  {v\Sigma_1\over [1+\Sigma(0)]^2}.
\end{equation}
For the case of one trap, $m=1$, the expressions for the effective
parameters clearly correspond 
to our earlier results of Eqs.~(\ref{eq:1trap-convection-gf-gaussian}) and
(\ref{eq:1trap-convection-Deff}).   Note that the precise
structure of the exponent and the prefactor in
Eq.~(\ref{eq:ntrap-convection-gf-max}) is different from those in
Eq.~(\ref{eq:1trap-convection-gf-gaussian}) which was obtained by a
more accurate calculation.

The effective diffusion approximation [Eq.~(\ref{eq:ntrap-convection-gf-max})]
is accurate for large $x$ near the maximum as long as the integral in
Eq.~(\ref{eq:ntrap-convection-gf}) remains dominated by the saddle-point in
Eq.~(\ref{eq:ntrap-convection-saddle}).  In particular, the poles of
response function (\ref{eq:ntrap-convection-response}) must be far from
$p_\star$.  This is easily satisfied in the case of ``shallow'' traps with
large release rates $B_i\gg |p_\star|$.

On the other hand, this condition could be simply violated in the presence of
``deep'' traps with relatively small $B_i$.  Over small time intervals 
compared to the typical dwell time $B_i^{-1}$, these traps may work in
the straining regime in which they would {\em not\/} contribute
to the effective dispersivity.  This situation may be manifested as an
apparent time-dependence of the effective drift velocity $v_0$ and/or
the dispersivity $\lambda_0$.

\subsection{Model with a continuous trap distribution}
\label{subsec:continuous}

The multitrap generalization given in Eq.~(\ref{eq:ntrap-convection}) for
filtration is clearly a step in the right direction if we want an accurate
description of the filtering experiments.

Indeed, apart from the special case of a regular array of identical
densely-packed spheres with highly polished surfaces, one expects the trapping
sites (e.g., the contact points of neighboring grains) to differ.  For small
particles such as viruses, even a relatively small variation in trapping energy
could result in a wide range of release rates $B_i$ differing by many orders
of magnitude\cite{Kim-2006,Yoon-2006}.  Under such circumstances, it is
appropriate to consider mean-field  models with continuous trap
distributions.

\begin{figure}[htp]
  \centering
    \includegraphics[width=.9\columnwidth]{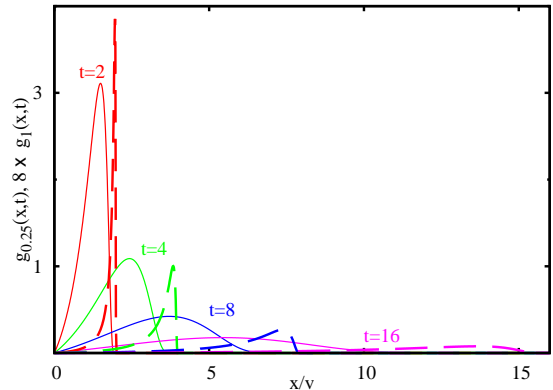}
    \caption{(Color online) Spatial dependence of the GF
      [Eq.~(\ref{eq:inftytrap-half-gf})] for the model presented in
      Eq.~(\ref{eq:ntrap-convection}) with continuous distribution of trap
      parameters corresponding to inverse-square-root singularity in the
      response function [see Eqs.~(\ref{eq:ntrap-convection-response}) and
      (\ref{eq:infty-convection-response-half}))].  Dashed lines show the GF at
      $\rho_{1/2}=0.25$, while solid lines present the same GF at
      $\rho_{1/2}=1$ multiplied by the factor of $8$.  We chose $t=2,4,8,12$
      as indicated in the plot.  Unlike in Fig.~\protect\ref{fig:comp}, due to
      abundance of traps with long release time, the GFs do not asymptotically
      converge toward a Gaussian form.}
  \label{fig:sqrt}
\end{figure}

Here we only consider a special case of a continuous distribution of
the trap parameters, $A_i$ and $B_i$, such that the release-rate density in
Eq.~(\ref{eq:ntrap-convection-response}) has an inverse-square-root
singularity, $\rho(B)=\rho_{1/2}/(\pi B^{1/2})$, with the release
rates ranging from infinity all the way to zero.  The corresponding
response function (\ref{eq:ntrap-convection-response}) could be expressed as
\begin{equation}
  \label{eq:infty-convection-response-half}
  \Sigma(p)=\rho_{1/2} /p^{1/2}.
\end{equation}
The inverse Laplace
transform [Eq.~(\ref{eq:ntrap-convection-gf})] gives the following GF:
\begin{equation}
  \label{eq:inftytrap-half-gf}
  g(x,t)=
\frac{ x\rho_{1/2}}{2 \sqrt{\pi }v \tau^{3/2}}
e^{-{x^2 \rho _{1/2}^2}/({4 v^2\tau})}\theta(\tau),
\quad \tau\equiv t-{x\over v}.
\end{equation}
Note that, in accordance with Eq.~(\ref{eq:ntrap-convection-gf-leading}),
there is no leading-edge $\delta$ function near $t=x/v$ as the expression for
the corresponding trapping rate $\beta$ diverges.  Because of the singular
behavior of $\Sigma(p)$ at $p=0$, there is no saddle-point expansion of the
form given in Eq.~(\ref{eq:ntrap-convection-saddle}).  Thus, there is no
Gaussian representation analogous to Eq.~(\ref{eq:ntrap-convection-gf-max}):
at large $t$, the maximum of the GF is located at $x_\mathrm{max}=v\rho_{1/2}
(2t)^{1/2}$, which is also of the order of the width of the Gaussian maximum.
The GF [Eq.~(\ref{eq:inftytrap-half-gf})] for two representative values of
$\rho_{1/2}$ is plotted in Fig.~\ref{fig:sqrt}.

We also note that for large $t$ at any given $x$, 
Eq.~(\ref{eq:inftytrap-half-gf}) has a power-law tail $\propto
t^{-3/2}$.   This property is generic for continuous trap density
distributions leading to small-$p$ power-law singularities in
$\Sigma(p)$.  For example, taking the density of the release rates as
a power law in $B$,
\begin{equation}
  \rho(B)={\sin(\pi s)\over \pi}{\rho_s\over
    B^s},\label{eq:inftytrap-dos}
\end{equation}
where $s$ is the corresponding exponent, $0<s<1$, we obtain
$\Sigma(p)= \rho_s p^{-s}$, and the large-$t$ asymptotic of the GF at
a fixed finite $x$ scales as 
\begin{equation}
  g(x,t)\propto t^{s-2}. \label{eq:power-scaling}  
\end{equation}
Such a power law is an essential feature of continuous
distribution~(\ref{eq:inftytrap-dos}) of the detachments rates; it cannot be
reproduced by a discrete set of rates $B_i$ which always produce an
\emph{exponential} tail.

\section{Filtration under unfavorable conditions}
\label{sec:full}
\subsection{Multitrap model with saturation}
The considered linearized filtration model presented by
Eq.~(\ref{eq:ntrap-convection}) can be used to analyze filtration of identical
particles in small concentrations and over limited time interval as long as
the trapped particles do not affect the filter performance.  However, unless
the model is used to simulate tracer particle dynamics in which no actual
trapping occurs, it is unlikely that the model remains valid as the number of
trapped particles grows.

Indeed, one expects that a trapped particle changes substantially the
probability for subsequent particles to be trapped in its vicinity.  Under
{\it favorable} filtering conditions characterized by filter ripening
\cite{OMelia-Ali-1978,Chiang-2004}, the probability of subsequent particle
trapping {\it increases} with time as the number of trapped particles $n_i$
grows.  On the other hand, under {\it unfavorable} filtering conditions, where
the Debye screening length is large compared to the trap size $\ell$, for
charged particles one expects trapping probabilities $A_i(n_i)$ to {\em
  decrease\/} with $n_i$.

If repulsive force between particles is large, we can assume that only one
particle is allowed to be captured in each trap.  Subsequently, a single trap
can be characterized by an attachment rate $A_i$ when it is empty and a
detachment rate $B_i$ when it is occupied, and the mean-field trapping/release
dynamics for a given group of trapping sites can be written as
\begin{equation}
  \label{eq:nonlin-trap}
  \dot n_i =C\, A_i (1-n_i)-B_i n_i.
\end{equation}
Note that this equation is non-linear because it contains the product of $C
n_i$.  

Previously, similar filtering dynamics was considered in a number of
publications (see Refs.~\cite{Bradford-2002} and~\cite{Bradford-2003} and
references therein).  In the present work, we allow for a possibility
of groups of traps differing by the rate parameters $A_i$ and $B_i$.  The
distribution of rate parameters can also be viewed as an analytical
alternative of the computer-based models describing a network of pores
of varying diameter\cite{Redner-2000,Kim-2006,shapiro-2007}.

Our mean-field transport model is completed by adding the kinetic
equation for the motion of free particles with concentration $C$,
\begin{equation}
  \label{eq:nonlin-convection}
\dot C  +vC'+\sum_{i=1}^m N_i \dot n_i=0, 
\end{equation}
which has the same form as the linearized
equations [Eq.~(\ref{eq:ntrap-convection})] considered in Sec.~\ref{subsec:continuous}.

We note that for shallow traps with large release rates $B_i$, the
non-linearity inherent in Eq.~(\ref{eq:nonlin-trap}) is not important for
sufficiently small suspended particle concentrations $C$.  Indeed, if $C$ is
independent of time, the solution of Eq.~(\ref{eq:nonlin-trap}) saturates at
\begin{equation}
  n_i(C)={CA_i\over B_i+CA_i}.
  \label{eq:nonlin-equilibrium}
\end{equation}
For  small free-particle concentration $C$, or for any $C$ and
large enough $B_i$, the trap population is small compared to 1, and the
non-linear term in Eq.~(\ref{eq:nonlin-trap}) can be ignored.

Therefore, as discussed in relation with the linearized multitrap
model [see Sec.~\ref{sec:linearized}A and
  Eq.~(\ref{eq:ntrap-convection})], the effect of shallow traps is to
introduce dispersivity of the arrival times of the particles on
different trajectories.  For this reason, we are free to drop the
dispersivity term [cf.~the CDE model, Eq.~(\ref{eq:convection-diffusion})],
and use a simpler convection-only model~(\ref{eq:nonlin-convection})
with several groups of traps with density $N_i$ per unit water volume, 
characterized by the relaxation parameters $A_i$ and $B_i$.

\subsection{General properties: Stable filtering front}
The constructed non-linear equations [Eqs.~(\ref{eq:nonlin-trap}) and
(\ref{eq:nonlin-convection})] describe complicated dynamics which is difficult
to understand in general.  Here, we introduce the front velocity, a parameter
that characterizes the speed of deterioration of the filtering capacity.  

Consider a semi-infinite filter, with the filtering medium initially clean,
and the concentration $C(0,t)=C_A$ of suspended particles at the inlet
constant.  After some time, the concentration of deposited particles near the
inlet reaches the dynamical equilibrium $n_i(C_A)$
[Eq.~(\ref{eq:nonlin-equilibrium})] and, on average, the particles will no
longer be deposited there.  At a given inlet concentration, the filtering
medium near the inlet is saturated with deposited particles.  On the other
hand, sufficiently far from the inlet, the filter is still clean.  On general
grounds, there should be some crossover between these two regions.

\begin{figure}  
\includegraphics[width=0.9\columnwidth]{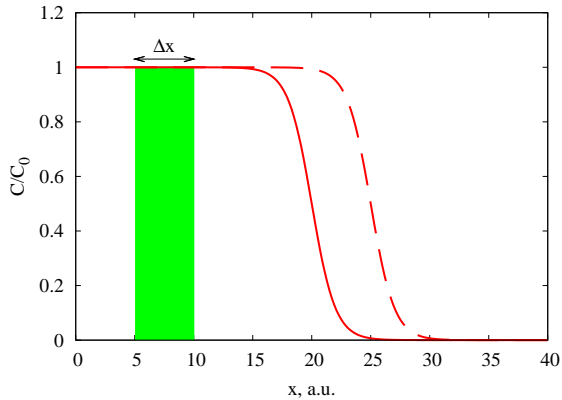}
\caption{Solid line shows the free particle concentration near a filtering
  front.  Dashed line shows the front shifted by $\Delta x$; the  additional
  free and trapped particles in the shaded region are 
  brought from the inlet [see Eq.~(\ref{eq:front-balance})].  See
  Eq.~(\ref{eq:front-C}) for exact front shape.}
\label{fig:ff}
\end{figure}

\begin{figure}  
\includegraphics[width=0.9\columnwidth]{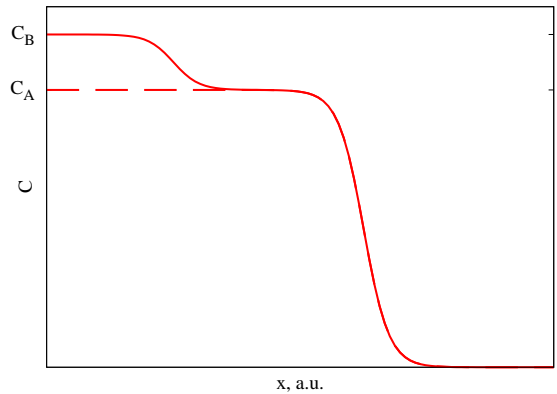}
\caption{Free particle concentration $C(x,t)$ with two filtering fronts.  The
  initial front moves on the background of clean filter and leaves behind the
  equilibrium filtering medium with $C=C_A$.  The secondary front with higher
  inlet concentration $C_B$ is moving on partially saturated medium.  With
  nonlinearity as in Eq.~(\ref{eq:nonlin-trap}), the secondary front is always
  faster, $v_{AB}>v_A$; the two fronts will eventually coalesce into a single
  front.}
\label{fig:twofront}
\end{figure}

The size of the saturated region grows with time [see
Fig.~\ref{fig:ff}].  The corresponding front velocity $v_A\equiv
v(C_A)$ can be easily calculated from the particle balance equation,
\begin{equation}
  \label{eq:front-balance}
  v_A C_A +v_A\sum_i N_i n_i(C_A)=  v C_A.
\end{equation}
This equation balances the number of additional particles needed to
increase the saturated region by $\delta x=v_A \delta t$ on the left,
with the number of particles brought from the inlet on the right [see
Fig.~\ref{fig:ff}].  The same equation can also be derived if we set
$C=C(x-v_A t)$, $n_i=n_i(x-v_A t)$ and integrate
Eq.~(\ref{eq:nonlin-convection}) over the entire crossover region.
The trapped particle density saturates as given by
Eq.~(\ref{eq:nonlin-equilibrium}), and the resulting front velocity is
\begin{equation}
  \label{eq:front-velocity-A}
  v(C_A)={v\over \displaystyle 1+\sum_i {N_i A_i \over A_i C_A+B_i}}. 
\end{equation}
This is a monotonously increasing function of $C_A$: larger inlet
concentration $C_A$ leads to higher front velocity, which implies that the
filtering front is stable with respect to perturbations.  Indeed, in
Appendix we show that the velocity $v_{AB}$ of a
secondary filtering front with the inlet concentration $C_B>C_A$ (see
Fig.~\ref{fig:twofront}), moving on the background of equilibrium
concentration of free particles $C_A$, is higher than $v_A$, i.e.,
$v_{AB}>v_A$.  Thus, if for some reason the original filtering front is split
into two parts, moving with the velocities $v_A$ and $v_{AB}$, the secondary
front will eventually catch up, restoring the overall front shape.

We emphasize that the existence of the stable filtering front is in sharp
contrast with the linearized filtering problem [see
  Eq.~(\ref{eq:ntrap-convection})], where the propagation velocity
$v_0$ [Eq.~(\ref{eq:ntrap-convection-saddle})] is independent of the
inlet concentration, and any structure is eventually washed out
dispersively (the width of long-time GF does not saturate with time).
Also, in the case of the filter ripening, the nonlinear term in
Eq.~(\ref{eq:nonlin-trap}) will be negative and thus would prohibit
the filtering front solutions due to the fact that the secondary
fronts move slower, $v_{AB}<v_A$.  The non-linear problem with
saturation is thus somewhat analogous to Korteweg-de Vries solitons
where the dispersion and nonlinearity compete to stabilize the
profile\cite{kdv,drazin-book}.

\subsection{Exactly solvable case}
\subsubsection{General solution}
Compared to the linear case presented in Sec.~\ref{sec:linearized}, the
physics behind the non-linear equations [Eqs.~(\ref{eq:nonlin-trap}) and
(\ref{eq:nonlin-convection})] is much more complicated.  However, the structure
of these equations immediately indicates that non-linearity reduces filtering
capacity because trapping sites could saturate in this model [see
Eq.~(\ref{eq:nonlin-equilibrium})].  While the relevant equations can also be
solved numerically, a thorough understanding of the filtering system,
especially with large or infinite number of traps, is difficult to achieve.

To gain some insight about the role of the different parameters in the
filtering process, we specifically focus on the non-linear models presented by
Eqs.~(\ref{eq:nonlin-trap}) and (\ref{eq:nonlin-convection}) which can be
rendered into a linear set of equations, very similar to the linear multitrap
model [Eq.~(\ref{eq:ntrap-convection})].  To this end, we consider the case
where all trapping sites have the same trapping cross sections, that is, all
$A_i=A$ in Eq.~(\ref{eq:nonlin-trap}).  If we introduce the time integral
\begin{equation}
  \label{eq:exact-u}
  u(x,t)\equiv \int_0^t C(x,t') dt', 
\end{equation}
then Eq.~(\ref{eq:nonlin-trap}) after a multiplication by $\exp Au$ can be
written as
\begin{equation}
  \label{eq:exact-trap-integration}
  {\partial_t}\left(n_i e^{A u}\right)
  + B_i \left( n_i e^{A u}\right) =  {\partial_t} \left(e^{A u}\right).
\end{equation}
Clearly, these are a set of linear equations,
\begin{equation}
  \label{eq:exact-trap-linearized}
\dot a_i
  + B_i a_i = \dot w, 
\end{equation}
with the following variables
\begin{equation}
  w\equiv w(x,t)=e^{Au},\quad a_i\equiv a_i(x,t) =n_i\,w.
\label{eq:exact-substitution}
\end{equation}
Note that Eq.~(\ref{eq:nonlin-convection})
can also be written as a set of linear equations in terms of these
variables.  If we
integrate~Eq.~(\ref{eq:nonlin-convection}) over time, we find
\begin{equation}
  \label{eq:exact-convection-integrated}
  \dot u +v u'+\sum_{i=1}^m N_i n_i=0, 
\end{equation}
where we assumed initially clean filter,
$C(x,0)=n_i(x,0)=0$. Considering that 
$\dot w= A \dot u \,w$ and $w'=A u'\,w$, we obtain 
\begin{equation}
  \label{eq:exact-convection-linearized}
   \dot w +v  w' +A\, \sum_{i=1}^m N_i  a_i=0.
\end{equation}

The main difference of the linear
Eqs.~(\ref{eq:exact-trap-linearized}) and
(\ref{eq:exact-convection-linearized}) from
Eqs.~(\ref{eq:ntrap-convection}) is in their initial and boundary conditions,
\begin{eqnarray}
  \label{eq:exact-initial-condition}
  w(x,0)&=&1,\quad a_i(x,0)=0,\\ 
  \label{eq:exact-boundary-condition}
  w(0,t)&=&e^{A u_0(t)},\quad u_0(t)\equiv 
  \int_0^t dt'\, C(0,t').
\end{eqnarray}
Note that with the time-independent concentration of the particles in
suspension at the inlet, i.e., $C(0,t)=C_0$, boundary condition~(\ref{eq:exact-boundary-condition}) gives a growing exponent,
\begin{equation}
w_0(t)\equiv w(0,t)=e^{ A C_0 t}.  
\label{eq:exact-boundary-exponent}
\end{equation}

The derived equations can be solved with
the use of the Laplace transformation.  Denoting $\tilde w\equiv \tilde
w(x,p)= \mathcal{L}_p\{w(t)\}$ and eliminating the Laplace-transformed trap
populations $\tilde n_i(x,p)\equiv \mathcal{L}_p\{n_i(x,t)\}$, we obtain
\begin{equation}
  \label{eq:exact-laplace-eqn}
  (p \tilde w -1 )\left[1+\Sigma(p)\right]+v \tilde w'=0,\quad 
  \Sigma(p)\equiv A\sum_i {N_i\over p+B_i}.
\end{equation}
The response function $\Sigma(p)$ is identical to that in
Eq.~(\ref{eq:ntrap-convection-response}), and for the case of continuous trap
distribution we can also introduce the effective density of traps,
$\rho(B)\equiv A\sum_i N_i \delta(B-B_i)$.  The solution of
Eq.~(\ref{eq:exact-laplace-eqn}) and the Laplace-transformed boundary
condition [Eq.~(\ref{eq:exact-boundary-condition})] becomes
\begin{equation}
  \label{eq:exact-laplace-solution}
  \tilde w={1\over p}+\left[\tilde w_0(p)-{1\over p}\right]
 e^{-[1+\Sigma(p)] px/v },
\end{equation}
where $\tilde w(0,p)=\tilde w_0(p)$. Employing the same notation as in
Eq.~(\ref{eq:ntrap-convection-gf}), the real-time solution of
Eqs.~(\ref{eq:exact-trap-linearized}) and
(\ref{eq:exact-convection-linearized}) with the boundary conditions [Eqs.
(\ref{eq:exact-initial-condition}) and (\ref{eq:exact-boundary-condition})] can
be written in quadratures,
\begin{equation}
  \label{eq:exact-formal-solution}
  w(x,t)=1+\int_0^t dt' \,\left[w_0(t-t')-1\right]\,g (x,t').
\end{equation}
The time-dependent concentration can be restored from here with the
help of logarithmic derivative, 
\begin{equation}
  C(x,t)={1\over A}{\partial \,\ln w(x,t)\over \partial
  t}. \label{eq:exact-derivative}   
\end{equation}
\subsubsection{Structure of the filtering front}
In the special case $C(0,t)=C_0={\rm const}$, the integrated
concentration [Eq.~(\ref{eq:exact-u})] is linear in time at the inlet,
$u_0(t)=C_0t$, and $w(0,t)$ grows exponentially [see
Eq.~(\ref{eq:exact-boundary-exponent})].  This exponent determines the
main contribution to the integral in
Eq.~(\ref{eq:exact-formal-solution}) for large $t$ and $x$.  Indeed, 
in this case we can rewrite Eq.~(\ref{eq:exact-formal-solution}) exactly as 
$w(x,t)=1+J(C_0)-J(0)$, where
\begin{equation}
  \label{eq:exact-two}
  J(C_0)\equiv  e^{AC_0t}\int_0^t dt'\,e^{-AC_0 t'} g (x,t').
\end{equation}
Note that $J(0)$ is proportional to the solution of the linearized
equations [Eq.~(\ref{eq:ntrap-convection})] with time-independent inlet
concentration $C(0,t)=\rm const$ [see
Eq.~(\ref{eq:tracer-convolution})].  The corresponding front is moving with the
velocity $v_0$ [Eq.~(\ref{eq:ntrap-convection-saddle})] and is widening over
time [Eqs.~(\ref{eq:ntrap-convection-gf-max}) and
(\ref{eq:ntrap-eff-params})].  Thus, for $x/v_0-t$ positive and sufficiently
large, this contribution to $w(x,t)$ is small and can be ignored.  In the
opposite limit of large negative $x/v_0-t$, $J(0)=1$, which exactly cancels
the first term in Eq.~(\ref{eq:exact-formal-solution}).

On the other hand, the term $J(C_0)$ grows exponentially large with time.  At
large enough $t$,  the integration limit can be extended to
infinity, and the integration in Eq.~(\ref{eq:exact-two}) becomes a Laplace
transformation, thus 
\begin{eqnarray}
  \nonumber
  w(x,t)&\approx& 1+e^{AC_0t}\int_0^{\infty} dt'\,e^{-AC_0 t'} g (x,t')\\
  &=&1+e^{p_0 t}\,e^{-[1+\Sigma(p_0)] p_0x /v},\quad p_0\equiv
  AC_0.\qquad 
  \label{eq:front-w}
\end{eqnarray}
This results in the following free-particle concentration [see
Eq.~(\ref{eq:exact-derivative})],
\begin{equation}
  \label{eq:front-C}
  C(x,t)={C_0 \over
  e^{[x/v(C_0)-t]AC_0}+1}, 
\end{equation}
and the occupation of the $i$th trap [Eqs.~(\ref{eq:exact-trap-linearized}) and
(\ref{eq:exact-substitution})],  
\begin{equation}
  \label{eq:front-ni}
  n_i(x,t)={A\over B_i +AC_0} C(x,t),
\end{equation}
with the front velocity 
\begin{equation}
  \label{eq:front-velocity-C0}
  v(C_0)\equiv {v\over 1+\Sigma(A C_0)}. 
\end{equation}
Note that this coincides exactly with the general case presented in
Eq.~(\ref{eq:front-velocity-A}) if we set all $A_i=A$.

\subsubsection{Filtering front formation}
The approximation in Eq.~(\ref{eq:front-w}) is valid in the vicinity of the
front, $|x/v(C_0)- t|\lesssim (AC_0)^{-1}$, as long as $x/v_0-t$ is positive
and large.  Since $v(C_0)>v_0=v(0)$, this implies 
\begin{equation}
  \label{eq:front-condition}
  x \left[{1\over v_0}-{1\over v(C_0)}\right]\gg {1\over A C_0},
\end{equation}
which provides an estimate of the distance from the outlet where the
front structure [Eqs.\ (\ref{eq:front-C}) and (\ref{eq:front-ni})] is formed.
The exactness of the obtained asymptotic front structure can be
verified directly by substituting the obtained profiles in
Eqs.~(\ref{eq:nonlin-trap}) and (\ref{eq:nonlin-convection}).

The exact expressions in Eqs.~(\ref{eq:exact-formal-solution}) and
(\ref{eq:exact-derivative}) for the free-particle concentration can be
integrated completely in some special cases.  Here we list two such results
and demonstrate the presence of striking similarities in the profiles $C(x,t)$
between different models, despite their very different rate distributions.
Furthermore, we show that the corresponding exact
solutions [Eq.~(\ref{eq:exact-derivative})] converge rapidly toward the general
filtering front [Eq.~(\ref{eq:front-C})].

\noindent {\bf Single-trap model with straining.}
In Sec.~\ref{sec:1trap-straining}, we found the explicit expression
[Eq.~(\ref{eq:1trap-straining-gf})] for the GF in the case of the linear model
for two types of trapping sites with rates $A_1$ and $B_1$ and permanent sites
with the capture rate $A_0$.  The resulting GF (with $A_1=A_0=A$ and $B_1=B$)
can be used in Eq.~(\ref{eq:exact-formal-solution}) to construct the solution
for the corresponding model with saturation,
\begin{eqnarray}
  \label{eq:nonlin-1trap-straining}
  & & \dot C+v C'+N_0 \dot n_0+N_1\dot n_1=0,\\
  & & \dot n_0=A C (1-n_0),\quad 
  \dot n_1=A C (1-n_1)-B_1 n_1.\quad
\end{eqnarray}

Let us consider the special case of the inlet concentration,
$C(0,t)=C_0\,\theta(T-t)\,\theta(t)$, 
constant over the interval $0<t<T$, and zero afterwards.  The function
$w_0(t)$ [see Eq.~(\ref{eq:exact-boundary-condition})] is, then
\begin{equation}
  w_0(t)=\exp [ A C_0 \min (t,T)], 
  \label{eq:exact-boundary-exponent-T}
\end{equation}
and the integration in Eq.~(\ref{eq:exact-formal-solution}) gives 
\begin{eqnarray}
  \label{eq:nonlin-1trap-solution}
  \lefteqn{w=1+e^{-\beta\xi} 
 \left[W(t)-e^{A C_0 T}W(t-T)\right],} & &   \\
\lefteqn{ W(t)\equiv
  \theta(t-\xi) \Bigl\{\left[e^{AC_0(t-\xi)}-1\right]} &&
\nonumber\\
& &\!\!\!   +\int^t_\xi\!\!d\tau\, 
  e^{-B_1(\tau-\xi)}\left[e^{AC_0(t-\tau)}
 -1\right]\frac{d}{d\tau}I_0(\zeta_\tau)\Bigr\} 
  ,\qquad \;
\end{eqnarray}
where $\xi\equiv x/v$ and $\zeta_\tau$ is given in
Eq.~(\ref{eq:1trap-straining-argument}).  The concentration of free particles,
$C(x,t)$, can be now obtained through Eq.~(\ref{eq:exact-derivative}).  The
step function $\theta(t-{x}/{v})$ included in $w$ indicates that it takes at
least $t=x/v$ for a particle to travel a distance $x$.

\begin{figure}[htbp]
  \centering
  \includegraphics[width=0.9\columnwidth]{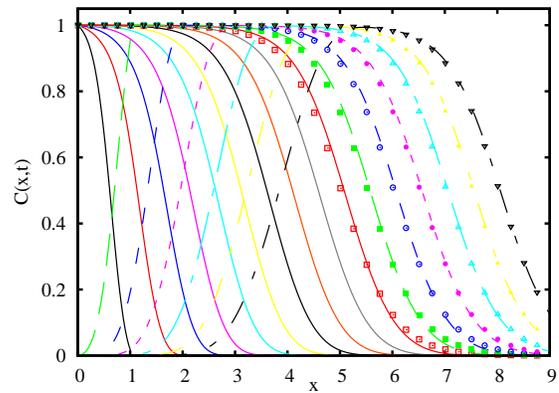}  
  \caption{(Color online) Formation of the filtering front for the
    single-trap filtering model with straining
    [Eq.~(\ref{eq:nonlin-1trap-straining})].  Lines show the
    free-particle concentration $C(x,t)$ extracted from
    Eq.~(\ref{eq:nonlin-1trap-straining}) with $T=10$,
    $A=v=C_0=1$, $N_0=0.388$, $N_1=3.60$, and $B_1=4.97$, for $t=1$, $2$, \ldots
    ,$16$. 
    Symbols show the front solution [Eq.~(\ref{eq:front-C})] for
    $t\ge 10$ with the front velocity [Eq.~(\ref{eq:front-velocity-C0})].}
  \label{fig:ots}
\end{figure}

\begin{figure}[htbp]
  \centering
  \includegraphics[width=0.9\columnwidth]{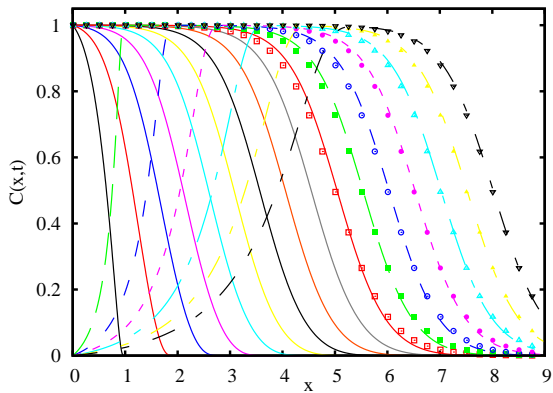}  
  \caption{(Color online) As in Fig.~\ref{fig:ots} but for filtering
    model~(\ref{eq:nonlin-convection}), Eq. (\ref{eq:nonlin-trap}) with continuous
    inverse-square-root trap distribution
    [Eq.~(\ref{eq:infty-convection-response-half})].  Parameters are
    $A=v=C_0=\rho_{1/2}=1$, $T=10$.  Symbols show the front solution
    [Eq.~(\ref{eq:front-C})] for $t\ge10$ with front velocity
    (\ref{eq:front-velocity-C0}).  The raising parts of the curves are almost
    identical with those in Fig.~\ref{fig:ots}, while there are some
    quantitative differences in the tails, consistent with the exponential vs\
    power-law long-time asymptotics of the corresponding solutions.}
  \label{fig:sqn}
\end{figure}

Figure \ref{fig:ots} illustrates $C(x,t)$ as a function of distance, $x$, at a
set of discrete values of time $t=1$, $2$, \ldots,$16$.  The model parameters
as indicated in the caption were obtained by fitting the response function
$\Sigma(p)=A N_0/p+AN_1/(p+B_1)$ at the interval $0.5<p<5.0$ to that of the
model with the continuous trap distribution (see Fig.~\ref{fig:sqn}).  The
solid lines show the curves for $t\le T$, while the dashed lines correspond to
$t>T$; they have a drop in the concentration near the origin consistent with
the boundary condition at the inlet.  The exact profiles show excellent
convergence toward the corresponding front profiles computed using
Eq.~(\ref{eq:front-C}) (symbols).

\noindent {\bf Model with square-root singularity.}
Let us now consider the non-linear model, [Eqs.~(\ref{eq:nonlin-convection})
and (\ref{eq:nonlin-trap})] with the inverse-square-root continuous trap
distribution, producing the response function given in
Eq.~(\ref{eq:infty-convection-response-half}).  The model is exactly solvable
if we set all $A_i=A$, while allowing the trap densities $N_i$ vary with $B$
appropriately.

The solution for the auxiliary function $w$ corresponding to the inlet
concentration $C(0,t)$ constant on an interval of duration $T$ is obtained by
combining Eqs.~(\ref{eq:exact-formal-solution}) and
(\ref{eq:exact-boundary-exponent-T}), with the relevant GF
[Eq.~(\ref{eq:inftytrap-half-gf})].  The resulting $x$-dependent curves $C(x,t)$
at a set of discrete time values are shown in Fig.~(\ref{fig:sqn}), along with
the corresponding asymptotic front profiles (symbols), for a parameter set as
indicated in the caption.  The solid lines show the curves for $t\le T$.  The
dashed lines are for $t>T$; they display a drop of the concentration near the
origin consistent with the boundary condition at the inlet.  Again, the
time-dependent profiles show gradual convergence toward front solution
(\ref{eq:front-C}).

Note that the profiles in Figs.~\ref{fig:ots} and \ref{fig:sqn} are
very similar even though the corresponding trap distributions differ
dramatically.  This illustrates that parameter fitting from a limited
set of breakthrough curves is a problem ill-defined mathematically.
The complexity and ambiguity of the problem grow with increasing
number of traps.  In Sec.~\ref{sec:experiment} we suggest an alternative
computationally simple procedure for parameter fitting using the data
from several breakthrough curves differing by the input
concentrations.

\section{Experimental implications}
\label{sec:experiment}

The suggested class of mean-field models is characterized by a large
number of parameters.  In the discrete case, these are the trap rate
constants $A_i$, $B_i$ and the corresponding concentrations $N_i$
along with the flow velocity $v$.  In the continuous case, the
filtering medium is characterized by the response function
$\Sigma(p)$ [see Eq.~(\ref{eq:ntrap-convection-response})].  In our
experience, two or three sets of traps are usually sufficient to
produce an excellent fit for a typical experimental breakthrough curve
(not shown).  This is not surprising, given the number of adjustable
parameters.  On the other hand, from Eq.~(\ref{eq:front-velocity-C0})
it is also clear that the obtained parameters would likely prove
inadequate if we change the inlet concentration.  The long-time
asymptotic form of the effluent during the washout stage would also
likely be off.

One alternative to a direct non-linear fitting is to use our result given in
Eq.~(\ref{eq:front-velocity-C0}) [or Eq.~(\ref{eq:front-velocity-A})] for the
filtering front velocity as a function of the inlet concentration, $C_0$.
With a relatively mild assumption that all trapping rates coincide, $A_i=A$,
one obtains the entire shape of the filtering front [Eq.~(\ref{eq:front-C})].
Thus, fitting the front profiles at different inlet concentrations $C_0$ to
determine the parameter $A$ and the front velocity $v(C_0)$ can be used to
directly measure the response function $\Sigma(p)$.

The suggested experimental procedure can be summarized as
follows.  ({\bf i}) One should use as long filtering columns as practically
possible in order to achieve the front formation for a wider range of
inlet concentrations.  ({\bf ii}) A set of breakthrough curves
$C(L,t)$ for several concentrations $C_0$ at the inlet should be
taken.  ({\bf iii}) For each curve, the front formation and the
applicability of the simplified model with all $A_i=A$ should be
verified by fitting with the front profile [Eq.~(\ref{eq:front-C})].  Given
the column length, each fit would result in the front velocity
$v(C_0)$, as well as the inverse front width $p=A C_0$.  ({\bf iv})
The resulting data points should be used to recover the functional
form of $\Sigma(p)$ and the solution for the full model. 

It is important to emphasize that the applicability of the model can be
controlled at essentially every step.  First, the time-dependence of each
curve should fit well with Eq.~(\ref{eq:front-C}).  Second, the values of the
trapping rate $A$ obtained from different curves should be close.  Third, the
computed washout curves should be compared with the experimentally obtained
breakthrough curves.  The obtained parameters, especially the details of
$\Sigma(p)$ for small $p$, can be further verified by repeating the
experiments on a shorter filtering column with the same medium.

\section{Conclusions}
\label{sec:conclusions}

In this paper, we presented a mean-field model to investigate the transport of
colloids in porous media.  The model corresponds to the filtration under
unfavorable conditions, where trapped particles tend to reduce the filtering
capacity, and can also be released back to the flow.  The situation should be
contrasted 
with favorable filtering conditions characterized by filter ripening. These
two different regimes can be achieved, e.g., by changing $p$H of the media if
the colloids are charged.  The unfavorable filtering conditions are typical
for filtering encountered in natural environment, e.g., ground water with
biologically active colloids such as viruses or bacteria.

The advantages of the model are twofold.  It not only fixes some
technical problems inherent in the mean-field models based on the CDE
but also admits analytical solutions with many groups of traps or
even with a continuous distribution of detachment rates.  It is the
existence of such analytical solutions that allowed us to formulate a
well-defined procedure for fitting the coefficients.  Ultimately, this
improves predictive capability and accuracy of the model.

The need for the attachment and detachment rate distributions under
unfavorable filtering conditions has already been recognized in the
field\cite{Bradford-2002,Bradford-2003,Yoon-2006}.  Previously it has
been implemented in computer-based models in terms of $ad$ $hoc$
distributions of the pore
radii\cite{Redner-2000,Kim-2006,shapiro-2007}.  Such models could
result in good fits to the experimental breakthrough curves.  However,
we showed in Sec.\ref{sec:experiment} that the relevant
experimental curves are often insensitive to the details of the trap
parameter distributions, especially on the early stages of filtering.

On the other hand, our analysis of the filtering front reveals that the front
velocity as a function of the inlet colloid concentration, $v(C_0)$
[Eq.~(\ref{eq:front-velocity-A})], is {\em primarily\/} determined by the
distribution of the attachment and detachment rates characterizing the filtering
medium.  We, indeed, suggest that the filtering front velocity is one of the
most important characteristics of the deep-bed filtration as it is directly
related to the loss of filtering capacity.

We have developed a detailed protocol to calculate the model parameters based
on the experimentally determined front velocity, $v(C_0)$.  We emphasize that
the most notable feature of the model is its ability to distinguish between
permanent traps (straining) and the traps with small but finite detachment
rate.  It is the latter traps that determine the long-time asymptotics of the
washout curves.

The suggested model is applicable to a wide range of problems in which
macromolecules, stable emulsion drops, or pathogenic micro-organisms such as
bacteria and viruses are transported in flow through a porous medium.  While
the model is purely phenomenological in nature, the mapping of the parameters
with the experimental data as a function of flow velocity and colloid size
will shed light on the nature of trapping for particular colloids.  The model
can also be extended to account for variations in attachment and detachment rates
for various colloids as needed to explain the steep deposition profiles near
the inlet of filters\cite{li-2004}.

\section*{ACKNOWLEDGMENT}

This research was supported in part under NSF Grants No.\ DMR-06-45668
(R.Z.), No.\ PHY05-51164 (R.Z.), and No.\ 0622242 (L.P.P.)

\appendix*

\section{Velocity of an intermediate front.}
\label{app:intermediate-front}
Here we derive an inequality for the velocity $v_{AB}$ of an intermediate front
interpolating between free-particle concentrations $C_A$ and $C_B$
[Fig.~\ref{fig:twofront}]. 

We first write the expressions for the filtering front velocities in clean
filter, with the inlet concentrations $C_A$ and $C_B>C_A$
[cf.~Eq.~(\ref{eq:front-balance})],  
\begin{eqnarray*}
\left({v\over v_A}-1\right)C_A & =& \sum_i N_i n_i(C_A),\\ 
\left({v\over v_B}-1\right)C_B & =& \sum_i N_i n_i(C_B).
\end{eqnarray*}
The velocity $v_{AB}$ of the filtering front interpolating between $C_A$ and
$C_B$ [Fig.~\ref{fig:twofront}] is given by
\begin{equation}
\left(  {v\over v_{AB}}-1\right)(C_B-C_A)  = \sum_i N_i [n_i(C_B)-n_i(C_A)].
\end{equation}
Combining these equations, we obtain 
\begin{equation}
  \label{eq:intermediate-front-v}
  {C_B\over v_B}-{C_A\over v_A}={C_B-C_A\over v_{AB}}.
\end{equation}
From here we conclude that the left-hand side (lhs) of Eq.~(\ref{eq:intermediate-front-v})
is 
positive.  Solving for $v_{AB}$ and expressing the difference $v_{AB}-v_A$, we
have
\begin{equation}
v_{AB}-v_A={C_B (v_B-v_A)\over \displaystyle\left({C_B\over v_B}-
{C_A\over v_A}\right) v_B}.
\end{equation}
For the model with saturation [Eq.~(\ref{eq:nonlin-trap})], we saw
that $v_B>v_A$, thus $v_{AB}>v_A$.



\begin{thebibliography}{10}
\bibitem{endnote}
Note that Eq.~(\ref{eq:1trap-straining-gf}) can also be written as a time derivative,
  $\displaystyle g(x,t)= e^{-\beta x/v-B_1(t- x/v)} {\partial\over \partial
    t}\Bigl\{ \theta(t-x/v) I_0(\zeta_t) \Bigr\}$.
\end{thebibliography}

\end{document}